\documentclass[twocolumn,aps,prd,showpacs,10pt,superscriptaddress]{revtex4-1}
\usepackage{amssymb}
\usepackage{amsmath}
\newcommand{\vect}[1]{\boldsymbol{#1}} % my vector symbol
\usepackage{multirow}
\usepackage{graphicx}
\usepackage[usenames,dvipsnames,svgnames]{xcolor}
\usepackage{hyperref}
\hypersetup{
pdfnewwindow=true,      % links in new window
colorlinks=true,        % false: boxed links; true: colored links
linkcolor=Blue,         % color of internal links (change box color with linkbordercolor)
citecolor=Blue,         % color of links to bibliography
filecolor=Blue,         % color of file links
urlcolor=Blue           % color of external links
}

\usepackage{soul}   %  st{strikethrough text}

\begin{document}
%\title{Interpretation of results in nonequilibrium molecular dynamics simulations of thermal transport and the importance of thermostatting details}
\title{Influence of Boundaries and Thermostatting on Nonequilibrium Molecular Dynamics Simulations of Heat Conduction in Solids}
\author{Zhen Li}
\affiliation{Jiangsu Key Laboratory of Advanced Food Manufacturing Equipment and Technology, Jiangnan University, 214122 Wuxi, China}
\author{Shiyun Xiong}
\email{s.y.xiong216@gmail.com}
\affiliation{Functional Nano and Soft Materials Laboratory (FUNSOM) and Collaborative Innovation Center of Suzhou Nano Science and Technology, Soochow University, 215123 Suzhou, P.R.China}
\author{Charles Sievers}
\affiliation{Department of Chemistry, University of California at Davis, One Shields Avenue, Davis, California 95616, USA}
\author{Yue Hu}
\affiliation{University of Michigan-Shanghai Jiao Tong University Joint Institute, Shanghai Jiao Tong University, Shanghai, China, 200240}
\author{Zheyong Fan}
\email{brucenju@gmail.com}
\affiliation{School of Mathematics and Physics, Bohai University, Jinzhou, China}
\affiliation{QTF Centre of Excellence, Department of Applied Physics, Aalto University, FI-00076 Aalto, Espoo,  Finland}
\author{Ning Wei}
\affiliation{Jiangsu Key Laboratory of Advanced Food Manufacturing Equipment and Technology, Jiangnan University, 214122 Wuxi, China}
\author{Hua Bao}
\email{hua.bao@sjtu.edu.cn}
\affiliation{University of Michigan-Shanghai Jiao Tong University Joint Institute, Shanghai Jiao Tong University, Shanghai, China,200240}
\author{Shunda Chen}
\affiliation{Department of Chemistry, University of California at Davis, One Shields Avenue, Davis, California 95616, USA}
\author{Davide Donadio}
\email{ddonadio@ucdavis.edu}
\affiliation{Department of Chemistry, University of California at Davis, One Shields Avenue, Davis, California 95616, USA}
\author{Tapio Ala-Nissila}
\affiliation{QTF Centre of Excellence, Department of Applied Physics, Aalto University, FI-00076 Aalto, Espoo, Finland}
\affiliation{Centre for Interdisciplinary Mathematical Modeling and Department of Mathematical Sciences, Loughborough University, Loughborough, Leicestershire LE11 3TU, UK}

\date{May 24, 2019}

\begin{abstract}
Nonequilibrium molecular dynamics (NEMD) has been extensively used to study thermal transport at various length scales in many materials. In this method, two local thermostats at different temperatures are used to generate a nonequilibrium steady state with a constant heat flux. Conventionally, the thermal conductivity of a finite system is calculated as the ratio between the heat flux and the temperature gradient extracted from the linear part of the temperature profile away from the local thermostats. Here we show that, with a proper choice of the thermostat, the nonlinear part of the temperature profile should actually not be excluded in thermal transport calculations. We compare NEMD results against those from the atomistic Green's function method in the ballistic regime, and those from the homogeneous nonequilibrium molecular dynamics method in the ballistic-to-diffusive regime. These comparisons suggest that in all the transport regimes, one should directly calculate the thermal conductance from the temperature difference between the heat source and sink and, if needed, convert it to the thermal conductivity by multiplying it with the system length. Furthermore, we find that the Langevin thermostat outperforms the Nos\'{e}-Hoover (chain) thermostat in NEMD simulations because of its stochastic and local nature. We show that this is particularly important for studying asymmetric carbon-based nanostructures, for which the Nos\'{e}-Hoover thermostat can produce artifacts leading to unphysical thermal rectification.
Our findings are important to obtain correct results from molecular dynamics simulations of nanoscale heat transport as the accuracy of the interatomic potentials is rapidly improving.
\end{abstract}

\maketitle

\section{Introduction}
Molecular dynamics (MD) is the most versatile and complete classical method to study heat transport at the nanoscale, which
is vital for many technological applications  \cite{cahill2014apr,volz2016epjp,donadio2018review} such as thermoelectric energy conversion and thermal management of electronic devices. As the interatomic interactions used in MD simulations have become inreasingly accurate by using 
quantum mechanical density functional based equilibrium MD \cite{Marcolongo:2015dn,carbogno2017prl,kang2017prb}, nonequilibrium MD \cite{stackhouse2010prl} and approach-to-equilibrium MD \cite{puligheddu2017prm,martin2018jncs}, it is crucial to develop a deeper understanding of the MD methods used for heat transport studies. In this work, we focus on one of the most popular MD methods for heat transport: the nonequilibrium MD (NEMD) method.

In NEMD simulations \cite{stackhouse2010rmg,shiomi2014arht}, one usually calculates the thermal conductivity $\kappa(L)$ of a finite system with length $L$ form the (presumably constant) temperature gradient $\nabla T$ and the heat flux $Q/S$ determined from a steady state according to Fourier's law:
\begin{equation}
\kappa(L) = \frac{Q}{S|\nabla T|}.    
\label{equation:kappa_NEMD}
\end{equation}
Here, $Q$ is the thermal power across a cross-section area $S$ perpendicular to the transport direction and $|\nabla T|$ is the magnitude of the temperature gradient. However, at the nanoscale transport is not \textit{diffusive} and the conventional concept of conductivity as a materials property becomes invalid \cite{datta1995}. 
For example, thermal transport at the nanoscale, especially in materials with high thermal conductivity such as graphene \cite{balandin2008nl,ghosh2008,Xu:2014gy}, is almost \textit{ballistic} with a length-dependent $\kappa(L)$ increasing  with increasing $L$. 
In this situation, a more appropriate quantity for describing heat transport is the thermal conductance per unit area $G(L)$, which is defined as
\begin{equation}
G(L) = \frac{Q}{S\Delta T}, 
\label{equation:G_NEMD}
\end{equation}
where $\Delta T > 0$ is the temperature difference between the heat source and the sink.
This quantity is constant in the ballistic regime and only weakly dependent on the system length in the nanoscale transport regime. 
In a system with a uniform cross section, the length-dependent conductivity and the conductance are related by the following equation:
\begin{equation}
G(L) \equiv \frac{\kappa(L)}{L}. 
\label{equation:G_and_kappa}
\end{equation}

A question then arises as to whether the conductivity and conductance calculated  using Eqs. (\ref{equation:kappa_NEMD}) and (\ref{equation:G_NEMD}) are consistent with each other. Clearly, if the temperature gradient is replaced by $\Delta T/L$, $\kappa$ and $G$ as calculated from Eqs. (\ref{equation:kappa_NEMD}) and (\ref{equation:G_NEMD}) imply Eq. (\ref{equation:G_and_kappa}). However, in most previous works using NEMD simulations, the temperature gradient was \textit{not} calculated as $\Delta T/L$, but was instead determined as the slope of the so-called linear region of the temperature profile, ignoring the nonlinear parts of the temperature profile near the thermal baths. This practice assumes that transport is diffusive, i.e. in accordance with Fourier's law that predicts a linear temperature profile at steady state conditions. It has been usually argued that nonlinearities are artifacts related to e.g., the local thermostats in the NEMD simulation setup and should be excluded. However, recent studies suggest that such nonlinearities have a physical origin, related to transport in finite size systems \cite{Allen2014prb,Cepellotti:2017km}. If the nonlinear parts need to be excluded, then $|\nabla T| \neq \Delta T/L$, leading to an inconsistency among  Eqs.~(\ref{equation:kappa_NEMD})-(\ref{equation:G_and_kappa}).

In this work we take graphene, a material with an exceptionally high lattice thermal conductivity and long phonon mean free paths \cite{balandin2008nl,ghosh2008}, as an example to explore the interpretation of NEMD results both in the ballistic and in the ballistic-to-diffusive transport regimes. In the ballistic limit, we compare NEMD (with spectral decomposition \cite{saaskilahti2014prb,saaskilahti2015prb,fan2017prb}) against the standard atomistic Green's function (AGF) method \cite{mingo2003prb,wang2008epjb,sadasivam2014arht}. We show that in order to reach an agreement with the AGF method, one needs to use Eq. (\ref{equation:G_NEMD}) to calculate the conductance and regard $\Delta T$ as the temperature difference between the hot and cold thermostats, not excluding any local nonlinear regions of the temperature profile. If one first calculates a conductivity using Eq. (\ref{equation:kappa_NEMD}) from the linear region of the temperature profile and then converts it to the conductance using Eq. (\ref{equation:G_and_kappa}), one obtains a ballistic conductance much larger than the correct one. 

On the other hand, although the NEMD method has been shown to be fully equivalent to the equilibrium MD (EMD) and the homogeneous nonequilibrium MD (HNEMD) methods in the diffusive limit \cite{fan2017prb,dong2018prb,fan2019prb,xu2018msmse,dong2018pccp,xu2019prb}, the influence of the simulation details in the NEMD method on the results in the ballistic-to-diffusive crossover regime have not been addressed. Here, we clarify the interpretation of the NEMD results by comparing them against those from the HNEMD method (also with spectral decomposition) \cite{fan2019prb}. Again, we show that one should not calculate the thermal conductivity using Eq. (\ref{equation:kappa_NEMD}) from a purely linear region of the temperature profile. Instead, the correct way is to calculate the conductance from Eq. (\ref{equation:G_NEMD}) with $\Delta T$ being the temperature difference between the heat source and sink and then convert it to the conductivity using Eq. (\ref{equation:G_and_kappa}). That is, one should calculate the conductivity as
\begin{equation}
    \label{equation:kappa_correct}
    \kappa(L) = \frac{Q}{S(\Delta T/L)}.
\end{equation}

Finally, we address thermal rectification in asymmetric graphene nanostructures, showing that the choice of thermostatting method and related parameters is of crucial importance, not only to calculate $G$  in the ballistic regime, but also to estimate correctly the thermal rectification efficiency of these systems.

This paper is organized as follows. In Sec. \ref{section:methods}, we review the various numerical phonon thermal transport methods used in this work, including the AGF method based on harmonic force constants (Sec. \ref{section:AGF}), the NEMD method (Sec. \ref{section:NEMD}), the HNEMD method (Sec. \ref{section:HNEMD}), and the spectral decomposition method (Sec. \ref{section:SHC}). After briefly presenting the AGF results in Sec. \ref{section:results_AGF}, we compare the NEMD results against the AGF results in Sec. \ref{section:results_NEMD_vs_AGF}. Then, we examine the temperature profiles in the NEMD simulations in Sec. \ref{section:results_temperature}. In Sec. \ref{section:results_diffusive}, we present the NEMD results for the  ballistic-to-diffusive transport regimes and compare them with the HNEMD results. Connection between the NEMD method and the Boltzmann transport equation method is discussed in Sec. \ref{section:results_BTE}. Finally in Sec.~\ref{section:rectification} we show the application of NEMD to thermal rectification in asymmetric graphene devices. In Sec. \ref{section:summary}, we give a summary and the main conclusions of this work.

\section{Methods\label{section:methods}}

\subsection{Atomistic Green's function method\label{section:AGF}}

Following the early work by Mingo and Yang \cite{mingo2003prb}, the AGF method has become a standard tool to study ballistic phonon transport in the harmonic approximation. In this method, one can calculate the phonon transmission $\mathcal{T}(\omega)$ between two leads as a function of the phonon frequency $\omega$. The ballistic conductance can be calculated through the Landauer expression as \cite{Imry:1999wo,Young:1989tm}:
\begin{equation}
    G=\int_0^{\infty} \frac{d\omega}{2\pi} G(\omega),
\end{equation}
where the spectral conductance $G(\omega)$ is 
\begin{equation}
    G(\omega)=G_{\rm c}(\omega)= \frac{k_{\rm B}}{S} \mathcal{T}(\omega)
\end{equation}
using classical statistics, and 
\begin{equation}
    G(\omega)=G_{\rm q}(\omega)= \frac{k_{\rm B}}{S} \frac{x^2e^x}{(e^x-1)^2}\mathcal{T}(\omega)
\end{equation}
using quantum statistics. Here, $x=\hbar\omega/k_{\rm B} T$, where $\hbar$ is Planck's constant, $k_{\rm B}$ is Boltzmann's constant, and $T$ is the system temperature.

As in the case of electron transport \cite{datta1995}, there are many equivalent representations of the phonon transmission \cite{wang2008epjb}. Here, we adopt Caroli's formula \cite{caroli2971jpc}:
\begin{equation}
    \mathcal{T}(\omega) = \rm{Tr}[\mathcal{G}(\omega)\Gamma_L(\omega) \mathcal{G}^{\dagger}(\omega) \Gamma_R(\omega) ],
\end{equation}
where 
\begin{equation}
    \mathcal{G}(\omega) = \frac{1}{\omega^2 - D - \Sigma_{\rm L}(\omega) - \Sigma_{\rm R}(\omega)}
\end{equation}
is the retarded Green's function for the system, $\mathcal{G}^{\dagger}(\omega)$ is the advanced Green's function, $D$ is the dynamical matrix of the system, and $\Sigma_{\rm L/R}(\omega)$ is the self energy matrix of the left (right) lead. The coupling matrices $\Gamma_{\rm L}(\omega)$ and $\Gamma_{\rm R}(\omega)$ are the imaginary part of the self energy matrices:
\begin{equation}
    \Gamma_{\rm L/R}(\omega) = i\left[\Sigma_{\rm L/R}(\omega) - \Sigma_{\rm L/R}^{\dagger}(\omega)\right].
\end{equation}
The dynamical matrix is the mass-weighted Hessian matrix of the empirical potential as used in our MD simulations:
\begin{equation}
    D_{\mu\nu} = \frac{1}{\sqrt{m_{\mu}m_{\nu}}} \frac{\partial^2 U}{\partial u_{\mu} \partial u_{\nu}},
\end{equation}
where $U$ is the total potential energy of the system and $u_{\mu}$ is a vibrational degree of freedom in the system with a mass of $m_{\mu}$. The self energy matrices can be obtained from the dynamical matrices for the semi-infinite leads by using an iterative method \cite{sancho1985jpf}. In numerical calculations, each atom contributes three  degrees of freedom. The elements of the dynamical matrix can also be expressed as
\begin{equation}
    D_{ij}^{ab} = \frac{1}{\sqrt{m_{i}m_{j}}} \frac{\partial^2 U}{\partial u_{i}^{a} \partial u_{j}^{b}},
\end{equation}
where $i$ and $j$ run over the atom indices, while $a$ and $b$ run over the three directions ($a,b=x,y,z$). From the relationship between force and potential, $F_i^a=-\partial U/\partial u_i^a$, we can write the above equation as 
\begin{equation}
    D_{ij}^{ab} = -\frac{1}{\sqrt{m_{i}m_{j}}} \frac{\partial F_i^a}{ \partial u_{j}^{b}}.
\end{equation}
This can be evaluated using a finite displacement $\delta$:
\begin{equation}
    D_{ij}^{ab} \approx \frac{1}{\sqrt{m_{i}m_{j}}} 
    \frac{F_i^a(r_j^b-\delta)-F_i^a(r_j^b+\delta)}{ 2\delta}.
\end{equation}
Here $F_i^a(r_j^b\pm\delta)$ is the total force on atom $i$ in the $a$ direction caused by displacing the position of atom $j$ in the $b$ direction by an amount of $\pm\delta$ (keeping all the other atoms at their relaxed positions). Here we use $\delta =0.005$~\AA.
%
%\DD{
%We note that the GF method is equivalent to the open-system lattice dynamics method proposed by Young and Maris~\cite{Young:1989tm} and relies on the same approximation of elastic scattering.}

\subsection{NEMD method\label{section:NEMD}}

There are many variants of the NEMD method, both in terms of boundary conditions in the transport direction and methods of generating the temperature difference. Periodic boundary conditions in the transport direction have been a popular choice, perhaps due to the ease of computer implementation. However, this choice cannot be directly compared to the AGF calculation setup, and is usually different from experimental situations, although it can give equivalent results as obtained using fixed boundary conditions as long as the sample length $L$ is carefully defined \cite{xu2018msmse}. We thus consider fixed boundary conditions in the transport direction. Regarding the methods of generating temperature difference, there are many algorithms, including the constant heat current method \cite{ikeshoji1994mp,jund1999prb}, the momentum swapping method \cite{muller-plathe1997jcp,Kuang:2010cm}, and methods based on thermostats. In both the constant heat current method and the momentum swapping method, the temperature difference cannot be precisely controlled. We therefore use thermostatting methods to generate the temperature difference.

\begin{figure}[htb]
\begin{center}
\includegraphics[width=8cm]{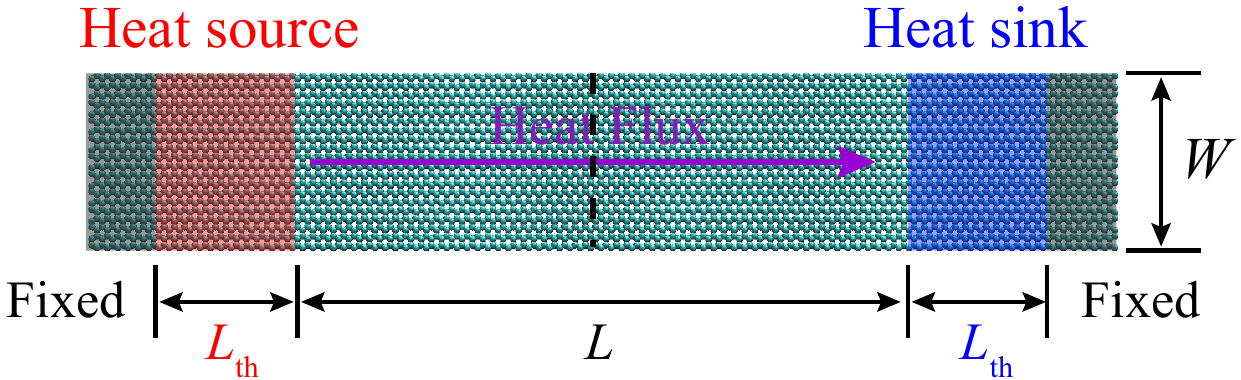}
\caption{Schematic illustration of the NEMD simulation setup. We consider a single-layer suspended graphene sheet with width $W$ and length $L$. A heat source region of length $L_{\rm th}$ is coupled to a thermostat with a temperature of $T+\Delta T/2$ and a heat sink region of the same length is coupled to a thermostat with a temperature of $T-\Delta T/2$. To prevent the atoms in the source and sink regions from sublimating and to keep the in-plane stress at zero, a few extra layers of atoms are fixed (forces and velocities are reset to zero during the time integration). Heat flux can be measured as the heat transfer rate $dE/dt$ in the local thermal baths divided by the cross-sectional area $S=Wh$, where $h=0.335$ nm is the convectional thickness of graphene. In this study, the width is fixed to $W=10$ nm and periodic boundary conditions are applied in the direction of $W$.}
\label{figure:model}
\end{center}
\end{figure}

In our NEMD simulations, we fixed a few layers of atoms at the two ends of the sample in the transport direction to achieve the fixed boundary conditions. Next to the two fixed layers, atoms within a length of $L_{\rm th}$ were coupled to a hot and a cold thermal bath, respectively. The distance between the two thermal baths defines the system length $L$, see Fig. \ref{figure:model} for an illustration. The whole simulation cell was first equilibrated at the target temperature using the Berendsen thermostat \cite{berendsen1984jcp} for $1$ ns, after which we switched off the global thermostat and switched on the local thermostats to realize the hot and cold thermal baths. The local thermostats were applied for $11$ ns, while steady state can be well achieved within the first $1$ ns in all our simulated systems. We used the data within the last $10$ ns to determine the temperature profile and the nonequilibrium heat current. Each NEMD simulation has been repeated three times and the statistical errors are very small and are thus omitted in the relevant figures.

To realize the local thermostats, we consider both the Nos\'{e}-Hoover chain \cite{nose1984jcp,hoover1985pra,martyna1992jcp} and the Langevin methods \cite{lepri_review_2003, dhar_review_2008}. In the Nos\'e-Hoover chain method , the equations of motion for the particles (with position $\vect{r}_i$, momentum $\vect{p}_i$, force $\vect{F}_i$, and mass $m_i$) in the thermostatted region (with $N_{\rm f}$ degrees of freedom) are 
%(those for the thermostat variables are not presented)
\begin{equation}
\frac{d \vect{r}_i}{dt} = \frac{\vect{p}_i}{m_i}; \quad
\frac{d \vect{p}_i}{dt} = \vect{F}_i - \frac{\pi_0}{Q_0} \vect{p}_i,
\end{equation}
where $Q_0=N_{\rm f} k_{\rm B} T \tau^2$ is the ``mass'' of the thermosat variable directly coupled to the system, with $\tau$ being a time parameter, and $\pi_0$ is the corresponding ``momentum''. In the Langevin method, the equations of motion for the particles in the thermostatted region are
\begin{equation}
\frac{d \vect{r}_i}{dt} = \frac{\vect{p}_i}{m_i}; \quad
\frac{d \vect{p}_i}{dt} = \vect{F}_i - \frac{\vect{p}_i}{\tau} + \vect{f}_i,
\label{equation:Langevin}
\end{equation}
where $\tau$ is a time parameter and $\vect{f}_i$ is a random force with a variation determined by the fluctuation-dissipation relation to recover the canonical ensemble distribution. 

There are many implementations of these thermostats and here we use a  velocity-Verlet integrator \cite{martyna1996mp,bussi2007pre}. In both thermostatting methods, there is a time parameter $\tau$ that dictates the coupling between the thermostat and the system. 
In the Nos\'{e}-Hoover chain method, the thermostat mass is proportional to $\tau^2$ and a larger $\tau$ would, in principle, decouple adiabatically the degrees of freedom of the thermostat from those of the system. However, when $\tau$ is too large the large inertia of the thermostat degrees of freedom produces unphysical fluctuations in the kinetic energy of the system \cite{Bussi:2007cs}.
In the Langevin method, $\tau$ is the inverse of the coefficient of friction and a larger $\tau$ also gives a weaker temperature control. The major difference between them is that in the Nos\'{e}-Hoover chain method, the temperature in the thermal bath region is adjusted globally and deterministically by re-scaling the velocities of the atoms with a common factor, while in Langevin dynamics, the velocities of the atoms are adjusted locally and stochastically.

\subsection{HNEMD method\label{section:HNEMD}}

Although the focus of this work is the NEMD method, we also use some results from the HNEMD method for comparison. This method is physically equivalent to the Green-Kubo method but is computationally much faster \cite{fan2019prb}. In this method, an external force of the form \cite{fan2019prb}
\begin{equation}
\vect{F}_{i}^{\rm ext}
= E_i \vect{F}_{\rm e} + \sum_{j \neq i} \left(\frac{\partial U_j}{\partial \vect{r}_{ji}} \otimes \vect{r}_{ij}\right) \cdot \vect{F}_{\rm e},
\end{equation}
is added to each atom $i$, driving the system out of equilibrium. Here, $E_i$ and $U_i$ are the total and potential energies of atom $i$, respectively, $\vect{r}_{ij}\equiv\vect{r}_{j}-\vect{r}_{i}$, and $\vect{r}_i$ is the position of particle $i$. The parameter $\vect{F}_{\rm e}$ is of the dimension of inverse length and should be small enough to keep the system within the linear response regime. A global thermostat should be applied to control the temperature of the system. The driving force will induce a
nonequilibrium heat current $\langle \vect{J} \rangle_{\rm ne}$ linearly proportional to $\vect{F}_{\rm e}$. For a given transport direction, this linear relation provides a way to compute the thermal conductivity in this direction:
\begin{equation}
\kappa(t) = \frac{\langle J(t)\rangle_{\rm ne}}{TVF_{\rm e}},
\label{equation:kappa_hnemd}
\end{equation}
where $T$ is the system temperature, $V$ is the system volume, $J=|\vect{J}|$, and $F_{\rm e}=|\vect{F}_{\rm e}|$. For a many-body potential, the heat current $\vect{J}$ is given by \cite{fan2015prb}
\begin{equation}
\label{equation:J}
\vect{J} =\sum_i \vect{v}_i E_i + \sum_i \sum_{j \neq i} \vect{r}_{ij}
\left(
\frac{\partial U_j}{\partial \vect{r}_{ji}} \cdot \vect{v}_i
\right),
\end{equation}
where $\vect{v}_i$ is the velocity of atom $i$. Because there is no boundary scattering in this method, the calculated thermal conductivity can be considered as that for an infinitely long system, as long as a sufficiently large periodic simulation cell is used. For graphene, a rectangular cell of dimension $25\times 25$ nm$^2$ is large enough to eliminate the finite-size effects \cite{fan2019prb}. 

\subsection{Spectral decomposition method \label{section:SHC}}

In both the NEMD and the HNEMD methods, a nonzero nonequilibrim heat current exists and can be spectrally decomposed. Considering an imaginary interface separating two groups of atoms $A$ and $B$ as schematically shown in Fig. \ref{figure:model}, this nonequilibrim heat current can be expressed as \cite{fan2017prb}
\begin{equation}
\label{equation:Q}
Q = - \sum_{i \in A} \sum_{j \in B}
\left\langle
\left(\frac{\partial U_i}{\partial \vect{r}_{ij}} \cdot \vect{v}_j
-\frac{\partial U_j}{\partial \vect{r}_{ji}} \cdot \vect{v}_i \right)
\right\rangle.
\end{equation}
For a spatially homogeneous system, $Q/S=J/V$, where $J$ is the magnitude of the heat current given in Eq. (\ref{equation:J}).

In the spectral decomposition method developed by S\"a\"askilahti \textit{et al.} \cite{saaskilahti2014prb,saaskilahti2015prb}, one first defines the force-velocity correlation function, which can be expressed as
\begin{equation}
\label{equation:Q_time}
K(t) = - \sum_{i \in A} \sum_{j \in B}
\left\langle
\left(\frac{\partial U_i}{\partial \vect{r}_{ij}} (0)\cdot \vect{v}_j(t)
-\frac{\partial U_j}{\partial \vect{r}_{ji}} (0) \cdot \vect{v}_i (t) \right)
\right\rangle
\end{equation}
for a general many-body potential \cite{fan2017prb}.
This correlation function reduces to the nonequilibrium heat current in Eq. (\ref{equation:Q}) at zero correlation time. The spectrally decomposed heat current $\tilde{K}(\omega)$ can be obtained from a Fourier transform of the force-velocity correlation function
\begin{equation}
\label{equation:K_omega}
\tilde{K}(\omega) = \int_{-\infty}^{\infty} dt e^{i\omega t}
K(t).
\end{equation}
The inverse Fourier transform is
\begin{equation}
K(t) = \int_{-\infty}^{\infty} \frac{d\omega}{2\pi} e^{-i\omega t} \tilde{K}(\omega).
\end{equation}
Because
\begin{equation}
Q = K(t=0)
= \int_{0}^{\infty} \frac{d\omega}{2\pi}
\left[2\tilde{K}(\omega)\right],
\end{equation}
we obtain the following spectral decomposition of the \textit{thermal conductance} in NEMD simulations:
\begin{equation}
G(\omega)
= \frac{2\tilde{K}(\omega)}{S\Delta T}
\quad
\text{with}
\quad
\label{equation:G_spectral}
G = \int_{0}^{\infty} \frac{d\omega}{2\pi} G(\omega).
\end{equation}
Similarly, the \textit{thermal conductivity} from HNEMD simulations can be spectrally decomposed as:
\begin{equation}
\label{equation:kappa_omega}
\kappa(\omega)
= \frac{2\tilde{K}(\omega)}{STF_{\rm e}}
\quad
\text{with}
\quad
\kappa = \int_{0}^{\infty} \frac{d\omega}{2\pi} \kappa(\omega).
\end{equation}

\subsection{Details on the numerical calculations\label{section:MD}}

The NEMD, HNEMD, and spectral decomposition methods have been implemented in the highly efficient Graphics Processing Units Molecular Dynamics (GPUMD) package \cite{fan2013cpc,fan2017cpc,gpumd}, which is the code we used for most of the MD simulations. The simulations of thermal rectification in asymmetric graphene devices were performed using the Large-scale Atomic/Molecular Massively Parallel Simulator (LAMMPS) package \cite{plimpton1995jcp}. We used the Tersoff potential \cite{tersoff1989prb} optimized for graphene systems \cite{lindsay2010prb}. For multilayer graphene the van der Waals interactions among different layers were modeled with the Lennard-Jones potential with $\varepsilon=3.296$~eV and $\sigma=3.55$~\AA ~\cite{Robertson:2015dp}.
A time step of $0.5$ fs, which ensures good energy conservation, was used in all the MD simulations. The AGF calculations were performed by using a Matlab code, which is also publicly available \cite{AGF-phonon-transport}. More details on simulations of thermal rectification are presented in Sec. \ref{section:rectification}.

\begin{figure}[htb]
\begin{center}
\includegraphics[width=8cm]{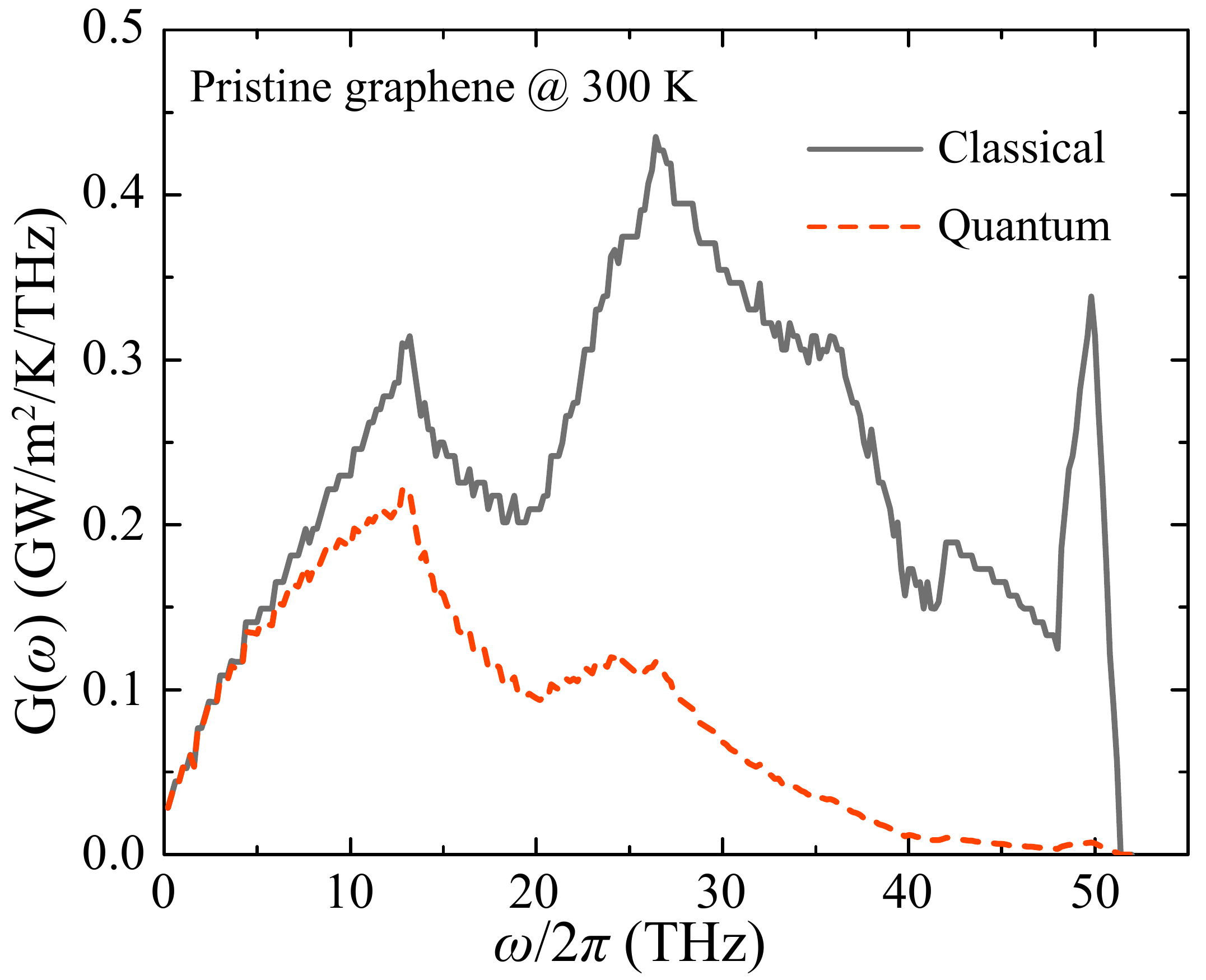}
\caption{Spectral ballistic thermal conductance as a function of phonon frequency obtained from the harmonic AGF calculations. The quantum thermal conductance is obtained from the classical one by multiplying a factor (related to spectral heat capacity) that is unity in the low-frequency limit and zero in the high-frequency limit.}
\label{figure:AGF}
\end{center}
\end{figure}

\section{Results and Discussion\label{section:results}}

\subsection{Ballistic thermal conductance from AGF calculations\label{section:results_AGF}}

We first calculate the ballistic conductance of the graphene sheet (with the same width as in the NEMD simulations) using the AGF method.  The ballistic spectral phonon conductance, obtained by using either classical or quantum statistics, is shown in Fig \ref{figure:AGF}. Because anharmonicity is totally absent in this method, the classical conductance (which is essentially the transmission) exhibits many quantized plateaus. By integrating the spectral conductance with respect to the frequency, we can get the total thermal conductance, which is about 12.2  and 4.2  GWm$^{-2}$K$^{-1}$ at room temperature in the cases of classical and quantum statistics, respectively. The quantum thermal conductance obtained here agrees with that reported by Serov \textit{et al.} \cite{serov2013apl} where the same empirical potential was used.

\begin{figure*}[htb]
\begin{center}
\includegraphics[width=14cm]{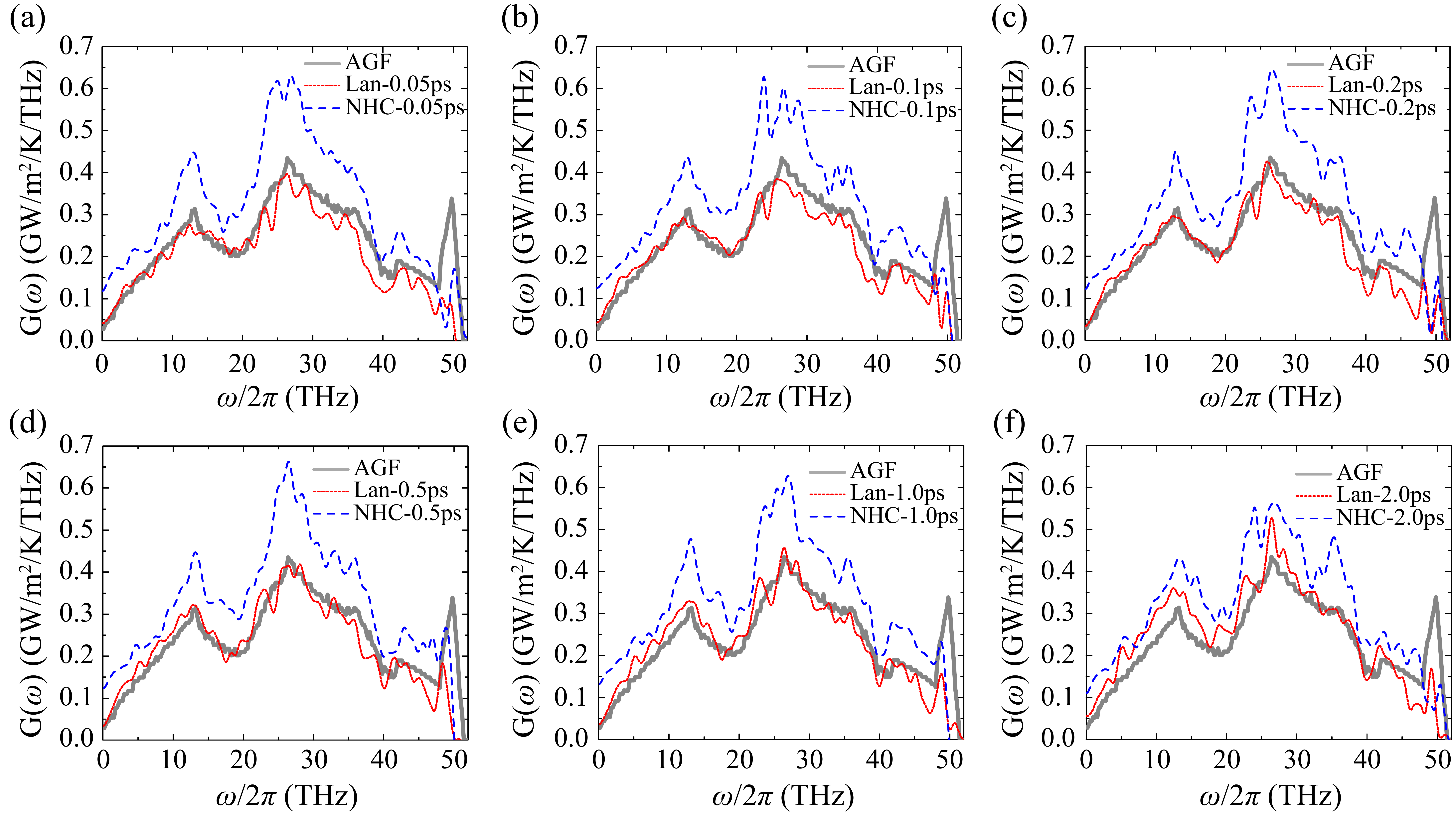}
\caption{Spectral conductance of a short graphene sheet with $L=10$ nm at 300 K from NEMD simulations using the Nos\'e-Hoover chain (thin dashed lines; labeled by NHC-$\tau$) and the Langevin (thin solid lines; labeled by Lan-$\tau$) thermostatting methods, compared to the fully ballistic conductance (thick solid lines) obtained from the harmonic AGF calculations. From (a) to (f), the time parameter $\tau$ in both thermostatting methods increases from  0.05 ps to 2 ps. The thermal baths are $L_{\rm th}=40$ nm long in all the simulations here.}
\label{f_tau}
\end{center}
\end{figure*}

\subsection{Comparison between NEMD and AGF in the ballistic regime\label{section:results_NEMD_vs_AGF}}

To obtain the ballistic conductance using the NEMD method, we set the system length as $L=10$ nm and the simulation temperature as $T=300$ K. The temperatures of the hot and cold thermal baths were set to $T\pm\Delta T/2$ with $\Delta T=10$ K. For such a short sample, anharmonicity caused by phonon-phonon scattering can be largely ignored (except for the optical phonons with the highest frequencies) and the phonon transport can be essentially regarded as ballistic. Here, we calculate the temperature difference from the actual average temperatures in the thermal baths. In the next subsection, we will examine the temperature profile and its implications in the interpretation of the results.

Figure \ref{f_tau} compares the spectral conductance obtained from the NEMD simulations (with spectral decomposition) using both the Nos\'e-Hoover chain and the Langevin thermostatting methods, against that from the AGF method. For the results from Fig. \ref{f_tau}(a) to \ref{f_tau}(f), the parameter $\tau$ in both thermostatting methods is increased from $0.05$ ps to $2$ ps. Because the MD simulations are classical, we need to compare with the classical ballistic thermal conductance from the AGF method.

The spectral conductance obtained by using the Nos\'e-Hoover chain thermostatting method is significantly larger than the reference AGF value, while that obtained by using the Langevin thermostatting method agrees with the AGF reference value well for all the $\tau$ values considered, particularly for intermediate $\tau$ values (0.1 ps to 1 ps). Chen {\it et al.} \cite{chen2010jpsj} also recommended using intermediate $\tau$ values in NEMD simulations of heat conduction. This is reasonable, as too small a $\tau$ results in too strong a perturbation on the system, and too large a $\tau$ results in too weak a control of the temperatures in the thermal baths.

\subsection{Temperature profile\label{section:results_temperature}}

\begin{figure*}[htb]
    \centering
    \includegraphics[width=14cm]{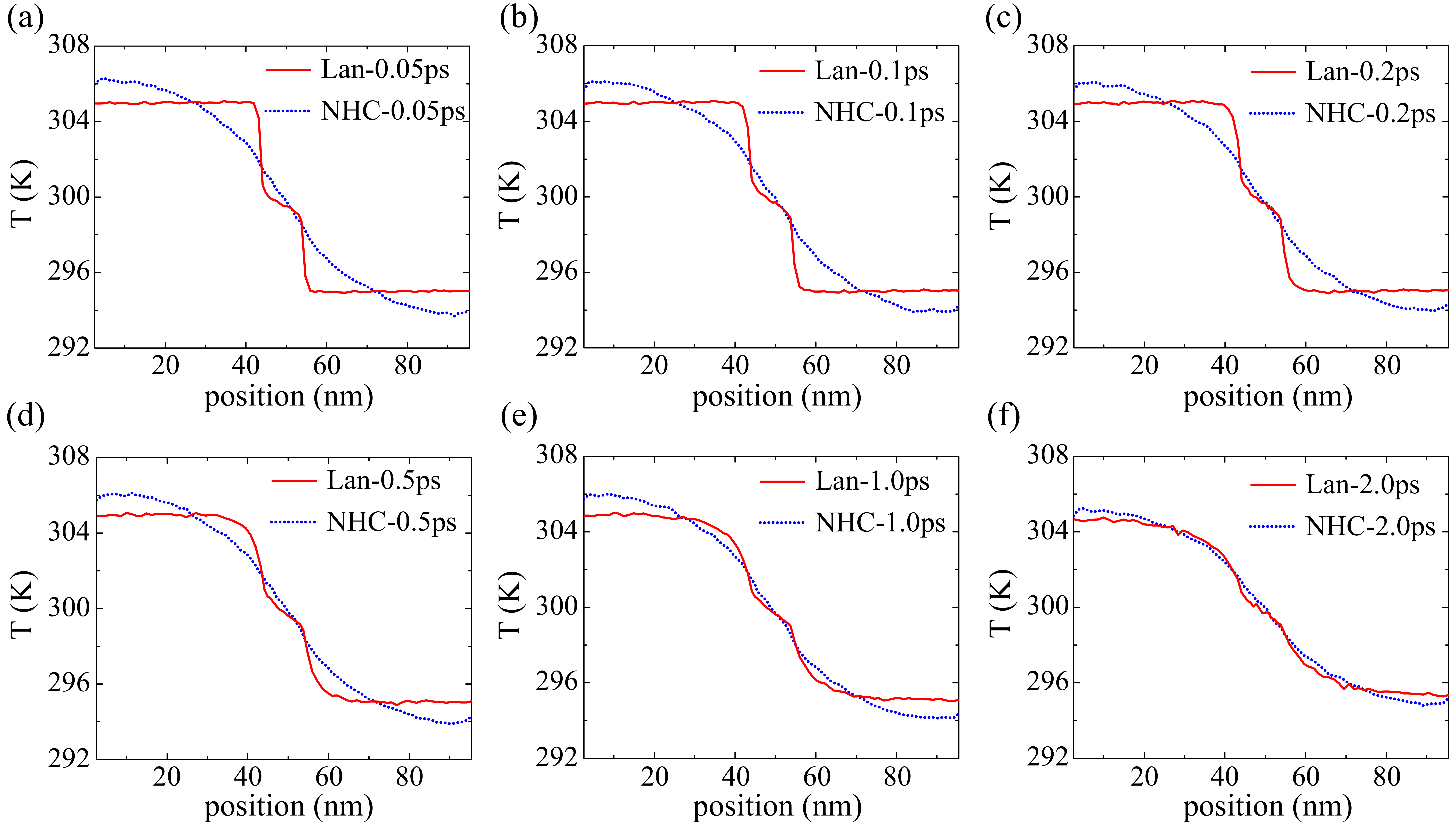}
    \caption{Temperature profiles in the NEMD simulations corresponding to the spectral conductances in Fig. \ref{f_tau}. From (a) to (f), the time parameter $\tau$ in both thermostatting methods increases from  0.05 ps to 2 ps. The thermal baths are $L_{\rm th}=40$ nm long in all the simulations here. In the legends, NHC-$\tau$ stands for Nos\'e-Hoover chain and Lan-$\tau$ stands for Langevin.}
    \label{figure:temp-ballistic-1}
\end{figure*}
\begin{figure*}[htb]
    \centering
    \includegraphics[width=14cm]{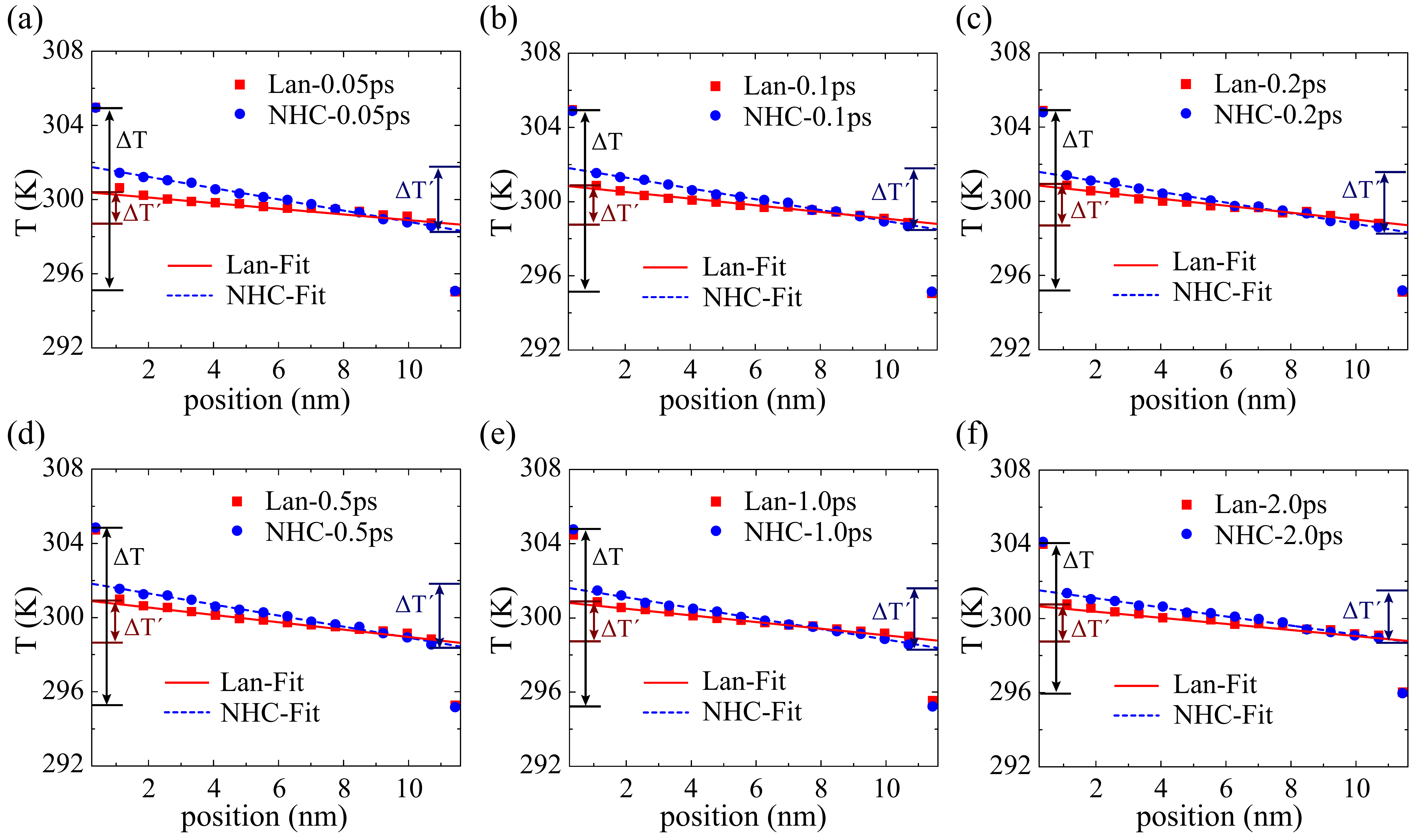}
    \caption{The temperature profiles from Fig. \ref{figure:temp-ballistic-1}, but with the thermal baths being treated as single points with zero length and the middle part of the system expanded in scale.}
    \label{figure:temp-ballistic-2}
\end{figure*}

To better understand the results above, we plot the detailed temperature profiles in the thermal baths as well as in the middle of the system with different coupling constants $\tau$ in Fig. \ref{figure:temp-ballistic-1}. Obviously, the Langevin thermostat gives much better temperature control than the Nos\'e-Hoover chain thermostat. When $\tau\leq 0.5$ ps, the temperatures in the thermal baths under the action of the Langevin thermostat are close to the target values. In contrast, the Nos\'e-Hoover chain thermostat cannot maintain a constant temperature in the thermal baths for any value of $\tau$. The temperatures close to the fixed boundaries overshoot the target source temperature or undershoot the target sink temperature. The reason is that Nos\'e-Hoover is a ``global" thermostat, i.e. it rescales the velocities of all the particles by the same amount at every MD step and it can guarantee only that the average kinetic temperature of the thermostatted region is the target one. In contrast Langevin is ``local" and it ensures that the whole thermostatted region is at the same temperature, as long as a strong coupling (small $\tau$) is enforced.
The use of Nos\'e-Hoover or any other global thermostat in NEMD results in an effective temperature difference that is larger than that calculated from the average temperatures in the thermal baths and an overestimated thermal conductance as shown in Fig. \ref{f_tau}. 

Let us now focus on the temperature profile in the middle of the system (the $L=10$ nm part). To this end, we represent the thermal baths as single points with the average temperatures within them; see Fig. \ref{figure:temp-ballistic-2}. We can see that there are abrupt temperature jumps between the thermal baths and the system in the middle. Previously, it has been frequently argued that one should apply Fourier's law to the linear region of the temperature profile only. This motivates to fit the middle part of the temperature profile using a linear function and extract a temperature gradient $|\nabla T'|=\Delta T'/L$, where $\Delta T'$ (c.f. Fig. 
\ref{figure:temp-ballistic-2}) is the temperature difference as determined by the interception between the fitted line for the middle part of the temperature profile and the vertical lines at the source and sink. The (effective) thermal conductivity for the system with length $L$ is then calculated as
\begin{equation}
\label{equation:k_prime}
    k'(L) = \frac{Q}{S(\Delta T'/L)}. 
\end{equation}
According to the relationship between the conductivity and conductance given in Eq. (\ref{equation:G_and_kappa}), this amounts to using the temperature difference $\Delta T'$ to calculate the thermal conductance:
\begin{equation}
\label{equation:G_prime}
    G'(L) = \frac{Q}{S\Delta T'}. 
\end{equation}

\begin{figure}[htb]
    \centering
    \includegraphics[width=8cm]{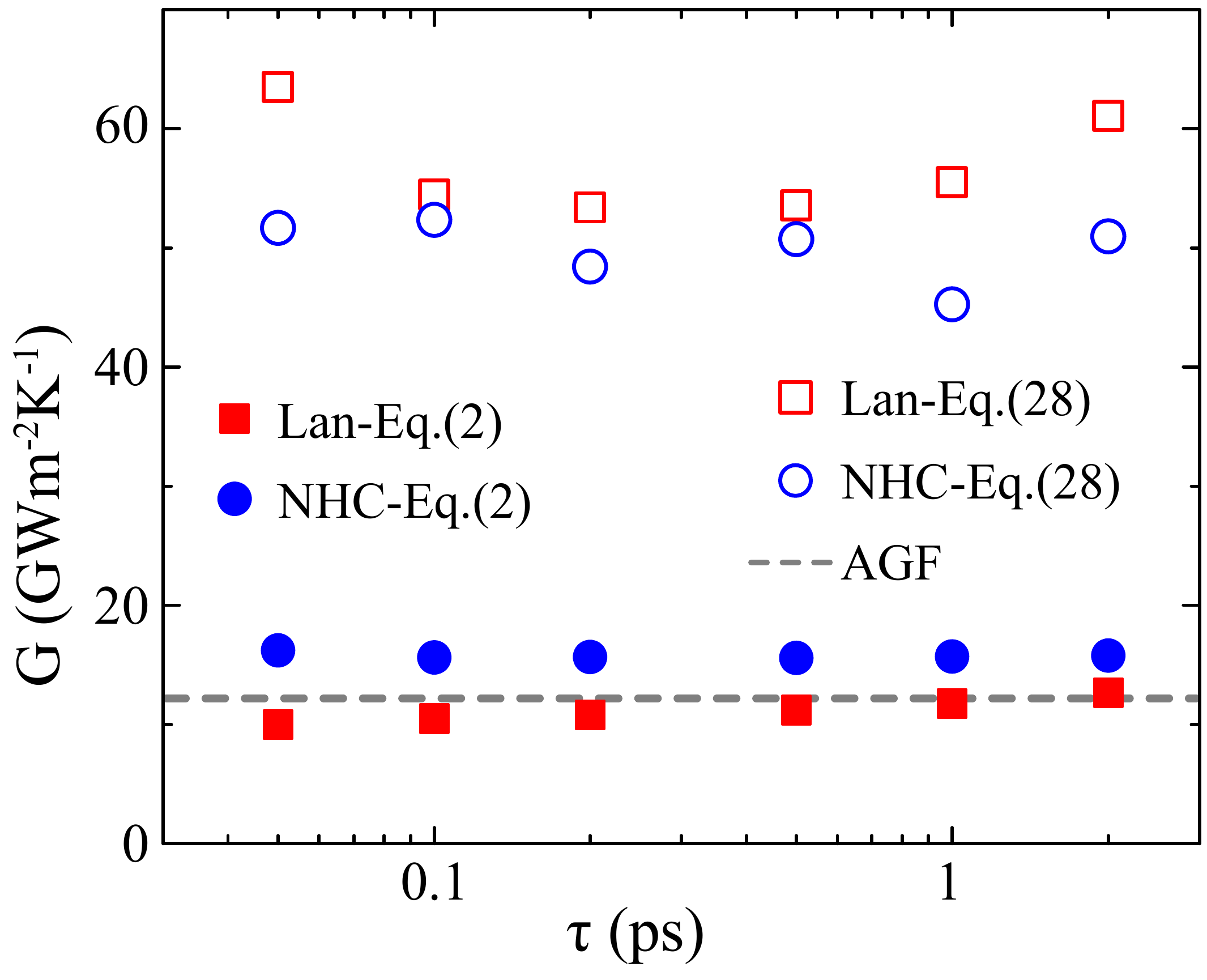}
    \caption{Total thermal conductance (per unit area) as a function of the relaxation time in the thermostating methods, obtained by using Eq. (\ref{equation:G_NEMD}) or Eq. (\ref{equation:G_prime}). The reference conductance value from AGF is represented as the dashed line.}
    \label{figure:G-and-G-prime}
\end{figure}

Figure \ref{figure:G-and-G-prime} shows the total thermal conductance (integrated over the frequency) values obtained by using Eqs. (\ref{equation:G_NEMD}) and (\ref{equation:G_prime}) and different thermostatting methods. When Eq. (\ref{equation:G_NEMD}) is used, the thermal conductance obtained with both thermostatting methods are relatively close to the AGF value, while the underestimation using the Langevin thermostat is due to the phonon-phonon scattering for the highest-frequency phonons, and the overestimation using the NHC thermostat is due to the nonuniform temperature in the source and sink regions. In contrast, using Eq. (\ref{equation:G_prime}), the obtained thermal conductance is several times higher than the AGF value, for both thermostatting methods. We thus reach the most important conclusion in this study: {\it Eqs. (\ref{equation:k_prime}) and (\ref{equation:G_prime}) are incorrect and should not be used}. The correct way is to calculate the thermal conductance using Eqs. (\ref{equation:G_NEMD}), taking $\Delta T$ as the temperature difference between the averaged temperatures in the source and sink. Correspondingly, the (effective) thermal conductivity should be calculated using Eq. (\ref{equation:kappa_correct}), as we will discuss in Sec. \ref{section:results_diffusive}.

\subsection{Comparison between NEMD and HNEMD in the ballistic-to-diffusive regime \label{section:results_diffusive}}

We now move from the ballistic regime to the ballistic-to-diffusive regime. To this end, we consider systems with different lengths: $L=25$, $50$, $100$, $200$, $500$, $1000$, and $2000$ nm. Other simulation parameters are fixed: $W=10$ nm, $\tau=0.2$ ps, $L_{\rm th} = 40$ nm. 
\begin{figure}[htb]
\begin{center}
\includegraphics[width=8cm]{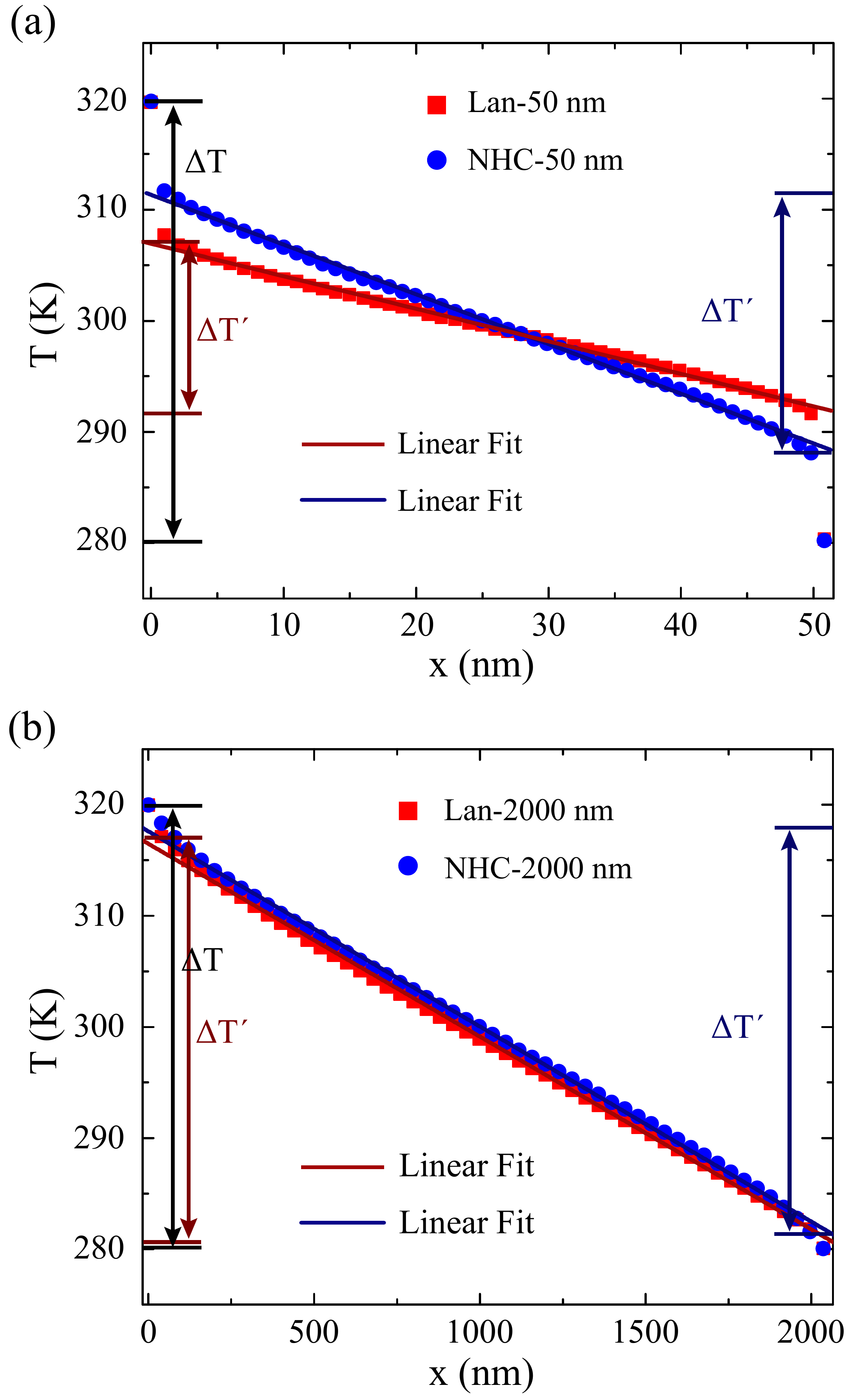}
\caption{Temperature profile in graphene sheets with two different lengths ($L=50$ and $2000$ nm) obtained by using the Langevin or the Nos\'e-Hoover chain thermostat. }
\label{figure:temp_diffusive}
\end{center}
\end{figure}
\begin{figure}[htb]
\begin{center}
\includegraphics[width=8cm]{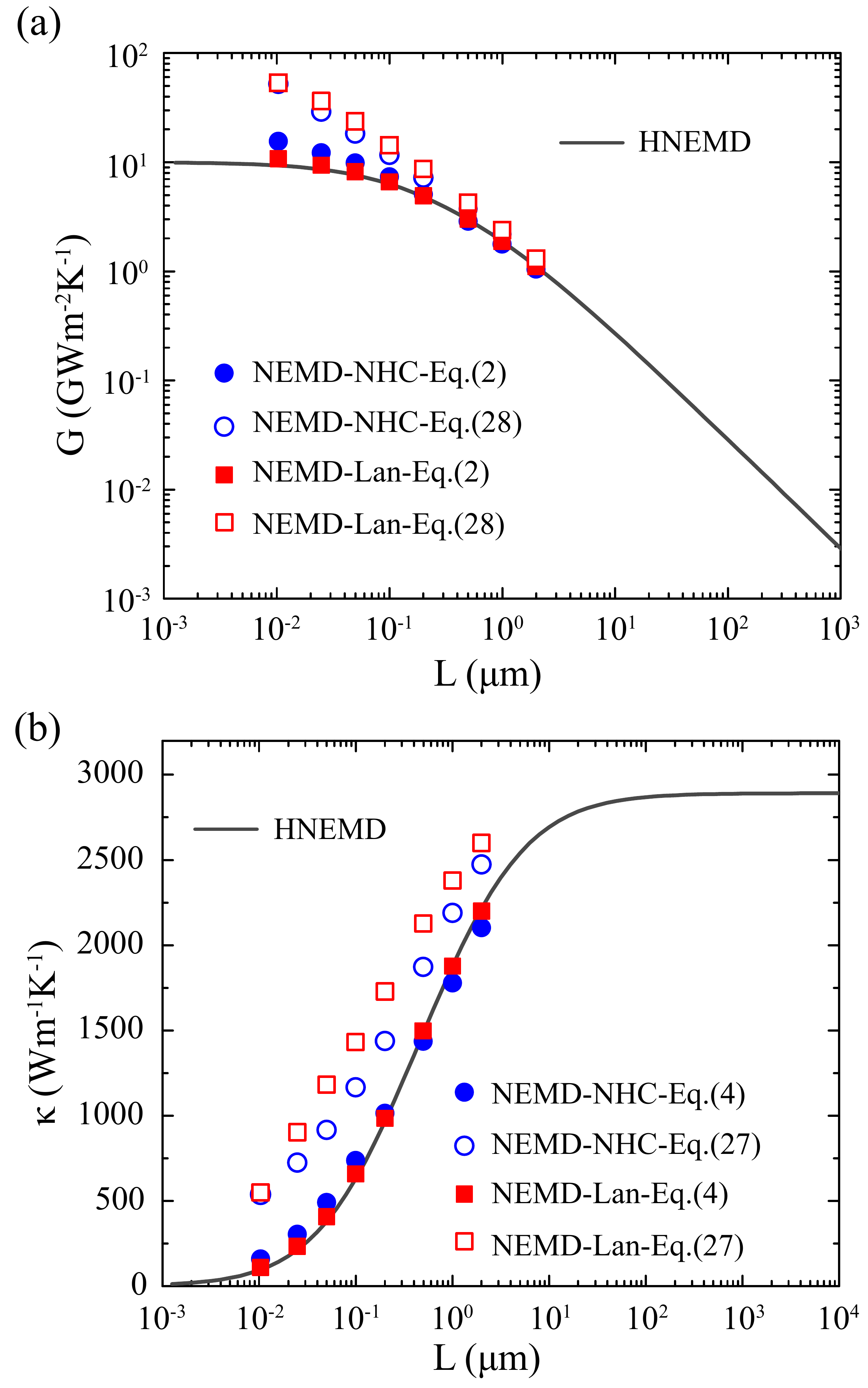}
\caption{(a) Thermal conductance and (b) thermal conductivity of graphene sheets as a function of system length from HNEMD and NEMD simulations. In the NEMD simulations, the conductance is  calculated by using either Eq. (\ref{equation:G_prime}) or Eq. (\ref{equation:G_NEMD}), and the conductivity is calculated by using either Eq. (\ref{equation:k_prime}) or Eq. (\ref{equation:kappa_correct}).}
\label{figure:diffusive-k}
\end{center}
\end{figure}

\begin{figure}[bht]
\begin{center}
\includegraphics[width=8cm]{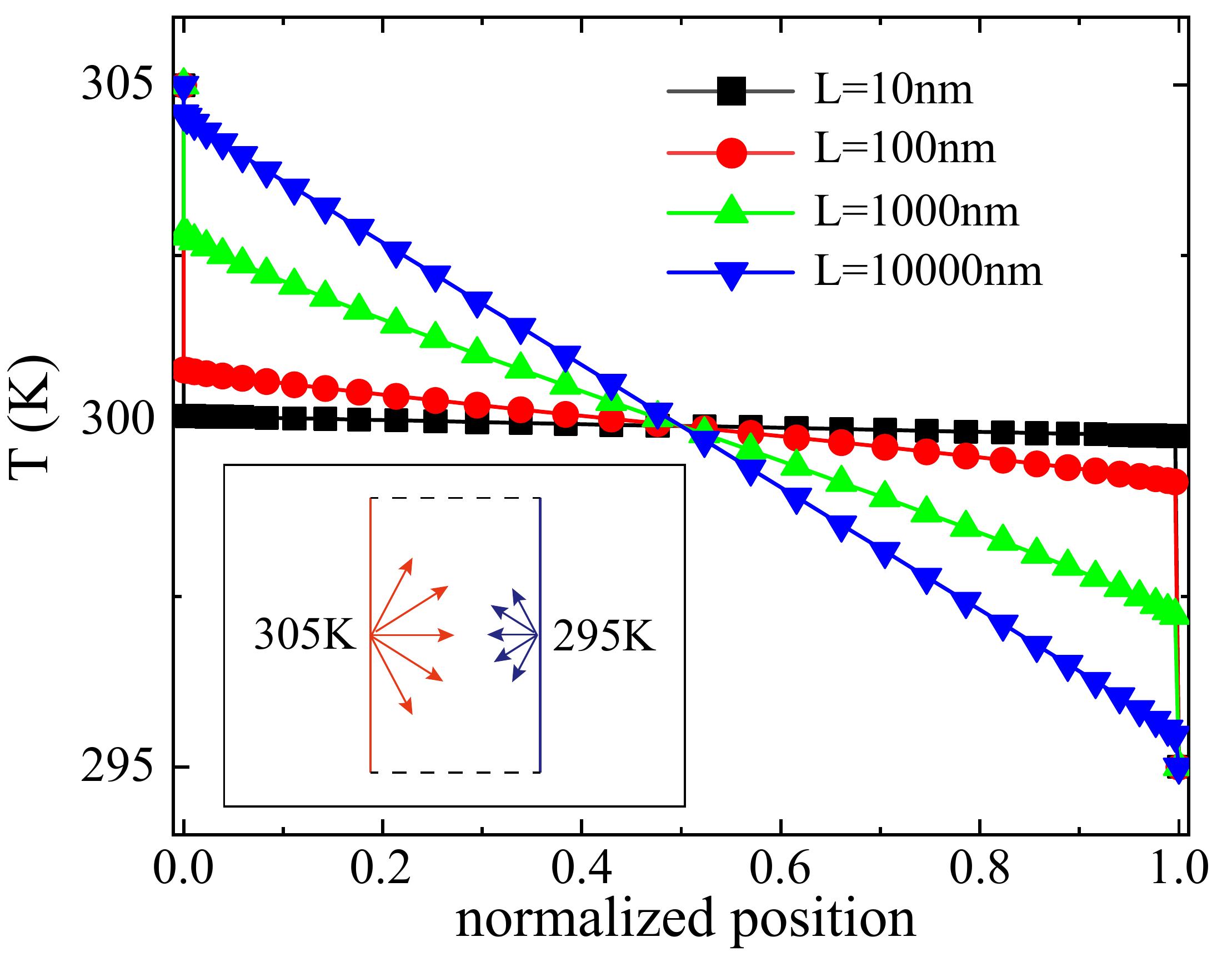}
\caption{Temperature profiles in graphene sheets with four different lengths ($L=10$, $100$, $1000$, and $10000$ nm) obtained by analytically solving a gray 1D BTE. The $x$ axis is the normalized position. The two boundaries are assumed to be at constant temperatures of 305 K and 295 K, respectively. Inset: a schematic illustration of the simulation domain and the thermalizing boundaries. }
\label{figure:BTE}
\end{center}
\end{figure}
Figure \ref{figure:temp_diffusive} shows the temperature profiles for two different system lengths obtained by using the Langevin or the Nos\'e-Hoover chain thermostat. The difference between the two definitions of temperature difference, $\Delta T$ and $\Delta T'$, decreases with increasing $L$, which is still significant when $L=50$ nm, but becomes much smaller when $L=2000$ nm. Due to this difference, the conductance as calculated from Eq. (\ref{equation:G_NEMD}) and that from Eq. (\ref{equation:G_prime}) are different, as can be seen from Fig. \ref{figure:diffusive-k}(a). From the comparison between the NEMD and AGF results in Sec. \ref{section:results_temperature}, we know that Eq. (\ref{equation:G_prime}) is wrong in the ballistic regime. To show that the corresponding expression for thermal conductivity Eq. (\ref{equation:k_prime}) is also wrong away from the ballistic regime, we convert the conductance in Fig. \ref{figure:diffusive-k}(a) into the conductivity as shown in Fig. \ref{figure:diffusive-k}(b) using Eq. (\ref{equation:G_and_kappa}).

There is a very efficient way to calculate the length-dependent thermal conductivity in the ballistic-to-diffusive regime based on the HNEMD-based spectral decomposition method developed recently \cite{fan2019prb}. In this method, one can calculate the spectral thermal conductivity $\kappa(\omega)$ using Eq. (\ref{equation:kappa_omega}). Then one can obtain the spectral phonon mean free path $\lambda(\omega)=\kappa(\omega)/G(\omega)$, from which one can calculate the length-dependent thermal conductivity as \cite{fan2019prb}:
\begin{equation}
\kappa(L) = \int_0^{\infty} \frac{d\omega}{2\pi} 
\frac{\kappa(\omega)}{1+\lambda(\omega)/L}.
\label{equation:kL}
\end{equation}
The length-dependent thermal conductivity calculated using this method as well as the corresponding thermal conductance [using Eq. (\ref{equation:G_and_kappa})] are shown as the solid lines in Fig. \ref{figure:diffusive-k}. One can see that the HNEMD results agree well with the NEMD values obtained by using Eqs. (\ref{equation:G_NEMD}) and (\ref{equation:kappa_correct}). The Langevin thermostat gives better agreement with the HNEMD results, but the difference between the results from the two thermostatting methods are quite small. The NEMD results obtained by using Eqs. (\ref{equation:G_prime}) and (\ref{equation:k_prime}), on the other hand, deviate from the HNEMD results significantly. The relative errors caused by using Eqs. (\ref{equation:G_prime}) and (\ref{equation:k_prime}) decrease with increasing system length $L$. If one focuses on the diffusive regime with relatively long systems, using Eq. (\ref{equation:k_prime}) will not result in large errors. This is why previous works \cite{fan2017prb,dong2018prb} using Eq. (\ref{equation:k_prime}) can get agreement between NEMD and other methods in the diffusive regime. However, when $L$ is relatively short (compared to the average phonon mean free path), using Eq. (\ref{equation:k_prime}) in NEMD simulations will result in large errors. 

\subsection{Relation to the Boltzmann transport equation  \label{section:results_BTE}}

The temperature drop between the thermal baths and the middle system is not an artifact of NEMD simulations, as it has also been observed in Boltzmann transport equation (BTE) calculations \cite{Allen2014prb}. To give a simple demonstration of such an effect, we analytically solved the gray 1D BTE for graphene, assuming an average phonon mean free path of 800 nm \cite{ghosh2008,Feng2015prb}. To be consistent with the NEMD (with the Langevin thermostat) and AGF simulations, the left and right boundaries are assumed to be at constant temperatures of 305 K and 295 K, respectively. 

The temperature profiles for different domain lengths $L$ are shown in Fig. \ref{figure:BTE}.  Note that in BTE the constant temperature (or thermalizing) boundary condition is implemented in the way that all the outgoing phonons leave the boundary unaffected, while all the incoming phonons have an intensity corresponding to equilibrium distribution at the given temperature \cite{saeid2018aa}, as schematically shown in the inset of Fig. \ref{figure:BTE}. Such an implementation is equivalent to having two boundaries in contact with infinitely large external thermal baths, and therefore is consistent with the Langevin thermostat as implemented in our NEMD simulations and also the AGF calculations. The temperature discontinuity is straightforward in the BTE picture. For example, in the ballistic limit, since phonons are not thermalized in the middle region, the phonons traveling from left to right have the same energy as the left boundary (305 K), while the phonons traveling from right to left have the same energy as the right boundary (295 K). Under such a non-equilibrium condition, the ``effective" temperature in the simulation domain is the average temperature of the two boundaries. Therefore, there is a temperature discontinuity at the boundary. As the system length increases, the phonons within the domain experience stronger phonon-phonon scatterings, so that the discontinuity gradually decreases and eventually diminishes. This is similar to what has been demonstrated in Fig. \ref{figure:temp_diffusive}. Note that for BTE simulation, it is well known that the conductance in this case should be calculated using the temperature difference between the boundaries \cite{chen2005nanoscale} instead of that in the middle region. Therefore, it further confirms that the conductance and conductivity should be calculated using Eq. (\ref{equation:G_NEMD}) and Eq. (\ref{equation:kappa_correct}), instead of Eq. (\ref{equation:G_prime}) and Eq. (\ref{equation:k_prime}). 

Note that here we have only considered a simplified gray BTE model to demonstrate that the temperature profiles in BTE calculations are qualitatively similar to those in NEMD simulations. Gu \textit{et al.} \cite{gu2019arxiv} have recently performed BTE calculations with a full iterative scheme, considering both three- and four-phonon scattering processes as well as boundary scatterings in graphene sheets. In such calculations, one can directly obtain length dependent thermal conductivity. They found that good agreement between NEMD and BTE regarding the length-dependence of thermal conductivity can only be obtained if Eq. (\ref{equation:kappa_correct}) is used to compute the thermal conductivity in NEMD simulations.  

\subsection{Thermal rectification in asymmetric graphene-based systems}
\label{section:rectification}

The discovery of thermal rectification in low-dimensional non-linear lattice models \cite{Terraneo:2002cz, li_ThermalDiode_2004} fostered significant efforts to identify efficient thermal diodes \cite{chen_ingredients_2015, benenti_thermalRectifier_2016, chen_efficient_2018}, which would lay the cornerstone of nanophononic circuitry~\cite{li2012rmp, volz2016epjp}.
In this context NEMD has been extensively used to probe thermal rectification in asymmetric graphene-based  nanodevices, including branched nanoribbons, triangular patches and multilayer junctions ~\cite{Hu:2009jj,Yang:2009gha,Zhong:2011bsa,Wang:2017hg,LopezSuarez:2018hd}. Using the definition of thermal conductance in Eq. (\ref{equation:G_NEMD}), the rectification factor is defined in terms of the difference between high and low thermal conductances $G_{\rm H}$ and $G_{\rm L}$  obtained by swapping the temperature bias of the thermal diode:
\begin{equation}
    \eta = \frac{G_{\rm H}-G_{\rm L}}{G_{\rm L}} \times 100\%.
\end{equation}
$G_{\rm H/L}$ may be replaced by the heat currents $J_{\rm H/L}$, provided that the temperature difference between the thermal reservoirs remains the same when the temperature bias is inverted. 
The above mentioned simulation works predict  extremely high $\eta$, up to 350$\%$, for carbon-based devices, but experimental measurements show much smaller thermal rectification, if any at all~\cite{chang_solid-state_2006, Wang:2017hg, Rojo:2019gr}.

Here we perform NEMD simulations of thermal rectification in both large trapezoid (LT) and small trapezoid (ST) monolayer graphene patches, where the large trapezoid compares to the smallest trapezoid studied in Ref. \onlinecite{Wang:2017hg}, which was reported to have large rectification, and in multilayer graphene junctions: bilayer to monolayer (BTM), trilayer to monolayer (TTM) and quadlayer to monolayer (QTM), where the multilayers are of the same geometry as those in \cite{Zhong:2011bsa}. All the layers were thermalized in order to compare with the previous numerical studies ~\cite{Zhong:2011bsa}. In all our thermal rectification simulations, we fixed two layers of atoms at the two  ends of the sample in the transport direction to achieve the fixed boundary conditions. Next to the two fixed layers, atoms within a length of $L_{\rm th}$ ($1.7$ nm for LT, $0.5$ nm for ST, and $0.8$ nm for other systems) were coupled to a hot and a cold thermal bath, respectively. We obtain the rectification factor of these nanodevices by computing the heat current in two separate NEMD simulations for each system, in which the hot and cold reservoirs are swapped, so to probe the high-conductance and the low-conductance conditions. 
We compare the rectification factor obtained using either  Nos\'{e}-Hoover, Nos\'{e}-Hoover chain, or Langevin dynamics to set the temperatures at 350 and 250 K for the hot and cold thermal reservoirs. 
After equilibration at 300 K, NEMD simulations are run for 4 ns (8 million time steps), and the thermal conductance is computed from the last 2 ns of these runs. 

 \begin{figure}[bt]
  \includegraphics[width=\linewidth]{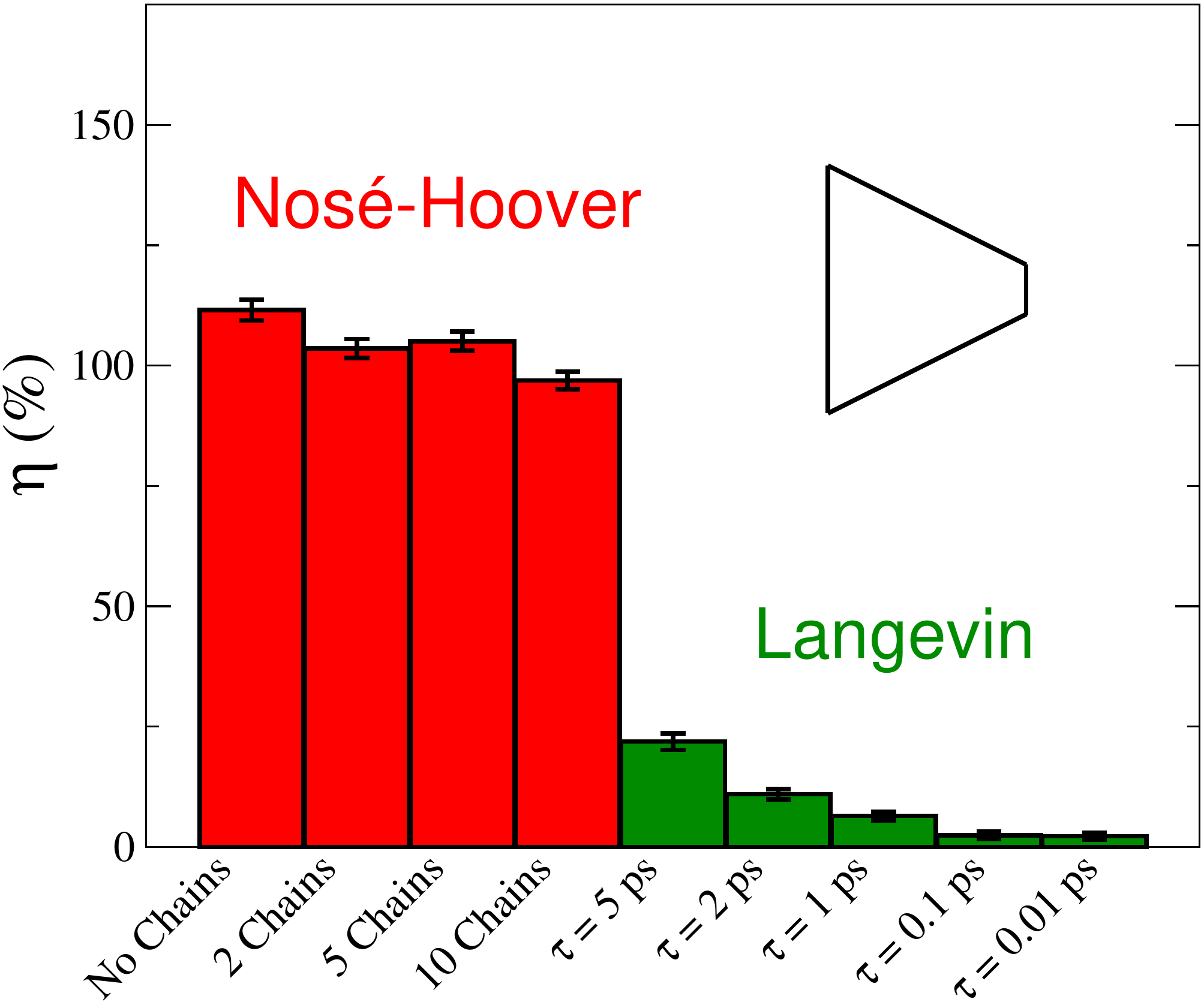}
  \caption{Comparison of the thermal rectification of the large trapezoid (LT) monolayer graphene obtained using either the Nos\'e-Hoover thermostat with different number of chains ($\tau=0.1$ ps)  or Langevin thermostat with different relaxation times.}
  \label{fig:thermostat_diode}
\end{figure}

We first test the effect of using different types of thermostats on the rectification factor of the LT graphene patch with 21.6 nm and 2 nm bases and height of 17 nm. This system is made of 8049 atoms and former NEMD simulations using the Nos\'{e}-Hoover thermostat predicted a rectification factor $\sim 95\%$ \cite{Wang:2017hg}.
Our simulations with the standard Nos\'{e}-Hoover thermostat with $\tau=0.1$ ps are in accord with these former results, predicting $\eta=111 (\pm 2)\%$. Adding more degrees of freedom to the thermostat, using Nos\'{e}-Hoover chains does not significantly change the estimate for $\eta$, which remains $\ge 100\%$ (Fig.~\ref{fig:thermostat_diode}). 
However, when Langevin dynamics is employed to fix the temperature of the thermal baths and produce a stationary flux, $\eta$ is considerably reduced which matches the results of a previous study \cite{chen2010jpsj}. Furthermore, $\eta$ depends on the relaxation time used in Eq. (\ref{equation:Langevin}). The smaller the $\tau$, i.e. the stronger the coupling, the lower the rectification factor and its uncertainty. For $\tau \le 0.1$ ps $\eta$ can be considered statistically zero.
As we have shown in the previous sections the NEMD Langevin dynamics provides more accurate results than Nos\'{e}-Hoover chain, and thus we can argue that the LT graphene patch considered here should not exhibit any significant thermal rectification. 

 \begin{figure}[hbt]
  \includegraphics[width=0.9\linewidth]{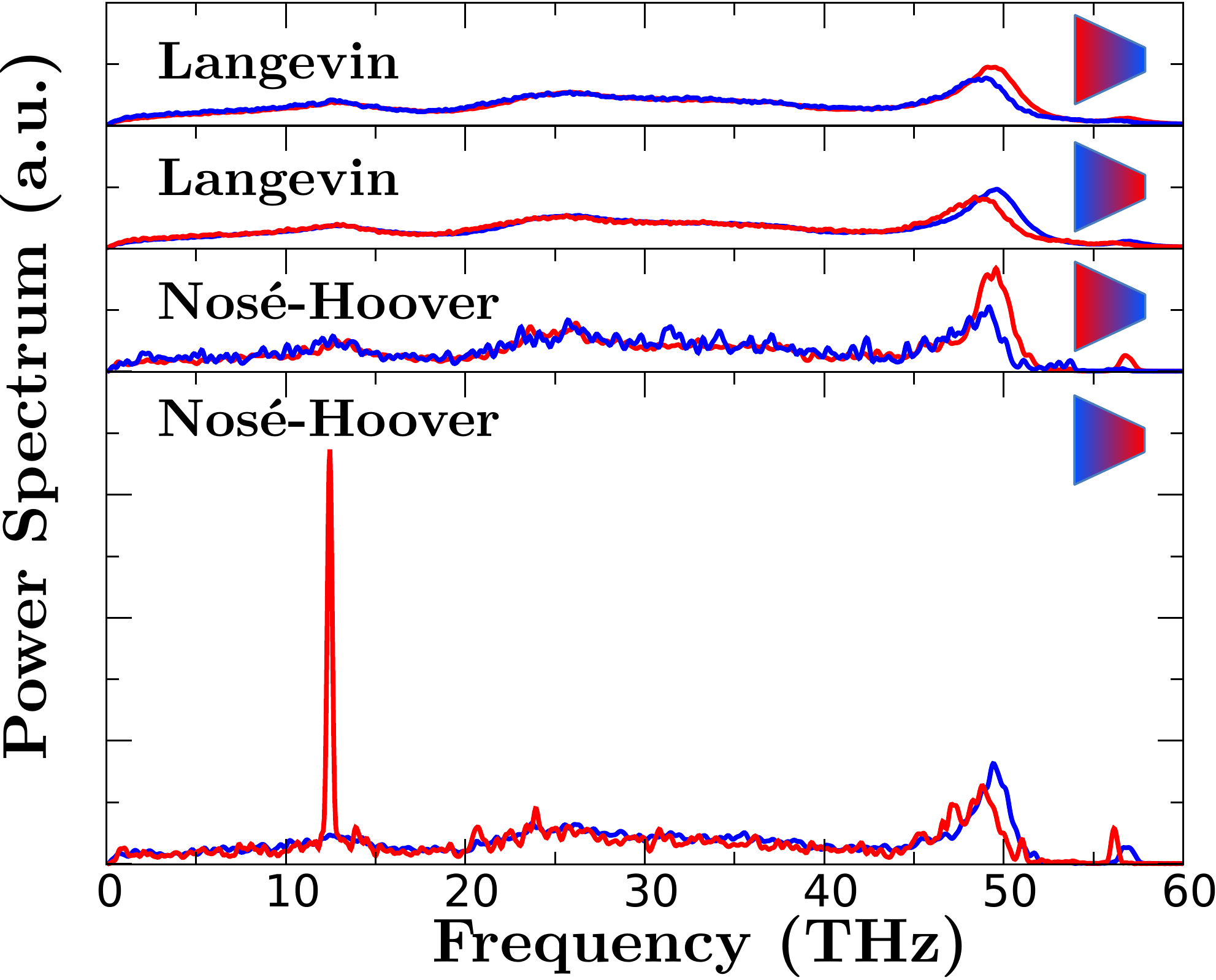}
  \caption{Power spectra of the thermal baths of the large trapezoid (LT) monolayer graphene system with different heat current biases using a Langevin thermostat with $\tau=0.1$ ps and a Nos\'e-Hoover thermostat with $\tau=0.1$ ps. The small insets show the geometry of the system with cold (blue) and hot (red) baths.}
  \label{fig:VDOSrectification}
\end{figure}

To unravel the origin of the large discrepancy between the two simulation methods, we calculated the power spectrum of the thermal baths in the NEMD simulations of trapezoid graphene at direct and reverse bias conditions (Fig.~\ref{fig:VDOSrectification}). When the hot bath is on the large base of the trapezoid the hot and cold power spectra overlap, except for the high frequency peak, which does not contribute significantly to heat transport. This condition corresponds to high conductance and the two thermostatting methods give similar results. 
With reverse bias, however, the hot bath exhibits an extremely intense peak at 12 THz in the case of the Nos\'{e}-Hoover chain thermostat, indicating that vibrational modes at this frequency are overpopulated. The energy accumulated in these specific modes cannot transfer through the device, leading to low conductance. Such overpopulation of a specific mode is an artifact of the Nos\'{e}-Hoover chain thermostat, occurring in non-ergodic system with a small number of degrees of freedom \cite{Patra:2014cp}. 
This effect leads to a disbalance between the power pumped by the thermostat into the hot bath and that removed from the cold bath, which should be equal at stationary conditions. With Nos\'{e}-Hoover chain these conditions are not achieved over the typical run time of several ns. 
When Langevin dynamics is used, the coupling parameter $\tau$ determines how rapidly stationary conditions with constant flux are attained, with shorter $\tau$ providing faster thermalization.

\begin{figure}[htb]
  \includegraphics[width=\linewidth]{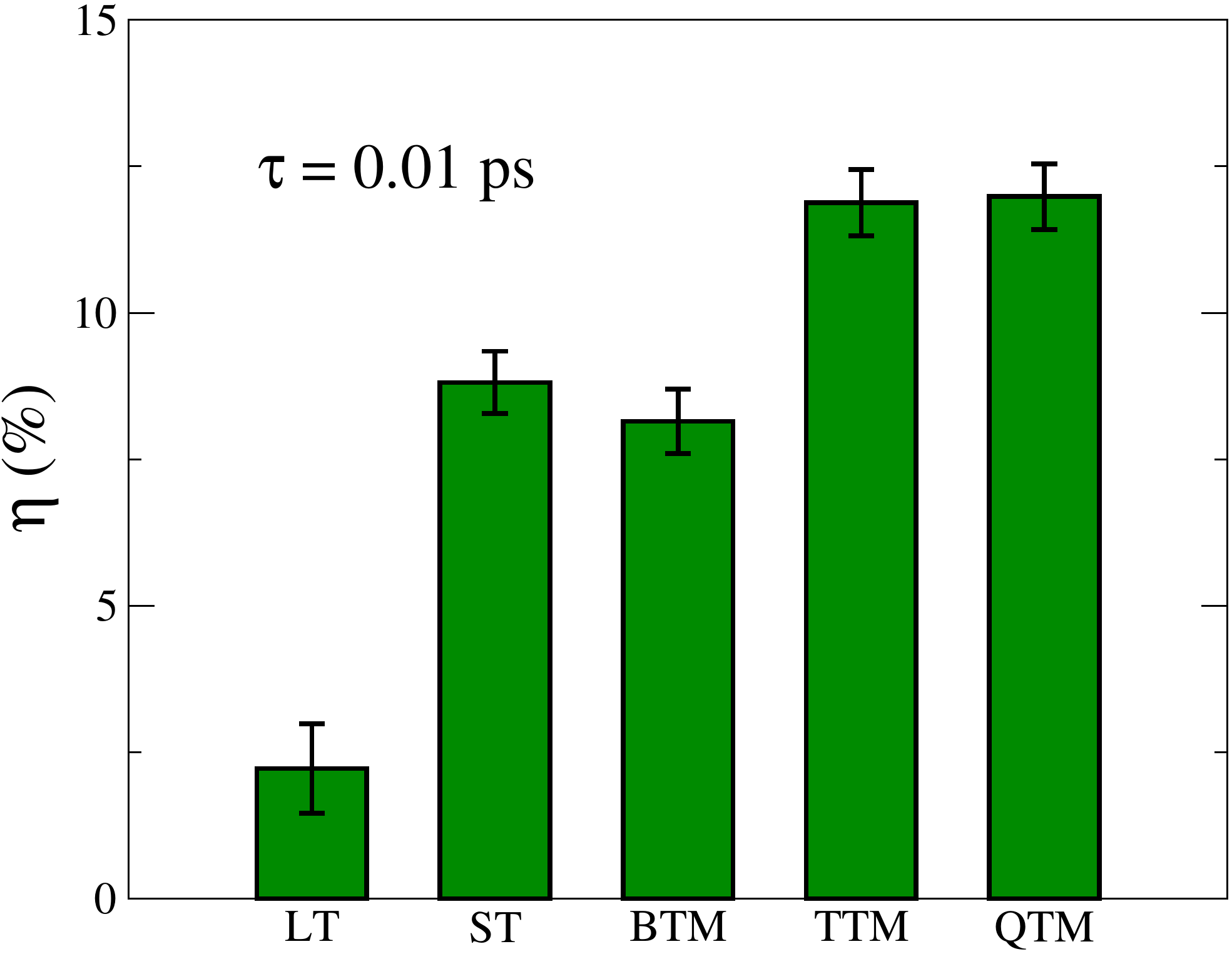}
  \caption{Rectification factor of large trapezoidal graphene monolayer (LT), small trapezoidal graphene monolayer (ST), and of multilayer graphene junctions, bilayer to monolayer (BTM), trilayer to monolayer (TTM), and quadlayer to monolayer (QTM).}
  \label{fig:device_recitification}
\end{figure}
%
%\DD{(say something about the difference between energy produced in  in NEMD?)}
%
These results suggest that Nos\'{e}-Hoover is unsuitable to study thermal rectification, unless one carefully addresses the very long thermalization time. Hence we use Langevin dynamics with $\tau=0.01$ ps to calculate the rectification factor of the other junctions considered. 
Results in Fig.~\ref{fig:device_recitification} show that for either trapezoidal or multilayer graphene junctions $\eta < 15\%$, in contrast with the high rectification efficiencies predicted in NEMD simulations carried out using Nos\'{e}-Hoover. 
Whereas multilayer graphene junctions may attain measurable rectification, all these systems are unsuitable for practical applications as thermal diodes. 

\section{Summary and conclusions\label{section:summary}}

In summary, we have carefully and systematically revisited some critical issues in the frequently used NEMD methods for thermal transport calculations. By comparing with the AGF method in the ballistic regime and the HNEMD method in the ballistic-to-diffusive regime, we found that the nonlinear part of the temperature profile in NEMD simulations should not be excluded in the calculations of the thermal conductivity and conductance. We also found that the Langevin thermostatting method controls the local temperatures better and is more reliable than the Nos\'e-Hoover (chain) thermostatting method for NEMD simulations. This is particularly important for studying asymmetric nanostructures, for which the Nos\'e-Hoover thermostat can produce artifacts leading to unphysical thermal rectification.
Based on our results, we recommend an intermediate value ($\tau =0.1-1$ ps) for the time parameter in the Langevin thermostat for thermal conductivity calculations and a small value ($\tau \leq 0.1$ ps) for thermal rectification studies, where one usually considers large temperature differences.

\begin{acknowledgments}
ZL and NW were supported by the National Natural Science Foundation of China (Grant No 11502217). SX acknowledges the support from the National Natural Science Foundation of China (Grant No. 11804242) and the Jiangsu Provincial Natural Science Foundation (Grant No. BK20160308). YH and HB acknowledge the support from the National Natural Science Foundation of China (Grant No. 51676121). ZF and TA-N acknowledge support from the Academy of Finland Centre of Excellence program QTF (Project 312298) and the computational resources provided by Aalto Science-IT project and Finland's IT Center for Science (CSC).  
\end{acknowledgments}

\end{document}